\documentstyle[12pt,titlepage,epsf,rotate]{article}
\def\_{\rule{.3em}{.15ex}} 

\newcommand{\scs}{\scriptscriptstyle}
\newcommand{\be}{\begin{equation}}
\newcommand{\ee}{\end{equation}}
\newcommand{\f}{\frac}
\newcommand{\ra}{\rightarrow}
\newcommand{\me}[1]{\langle#1\rangle}
\newcommand{\al}{\alpha_s}

\newcommand{\bsg}{$b \ra s \gamma$ }
\newcommand{\Bsg}{$B \ra X_s \gamma$ }
\begin{document}
\begin{titlepage}

 \begin{flushright}
  {\bf MPI-Ph/93-77\\
    TUM-T31-50/93}\\
    November 1993
 \end{flushright}

 \begin{center}
  \vspace{0.75in}

\setlength {\baselineskip}{0.3in}
  {\bf \LARGE Theoretical Uncertainties and Phenomenological Aspects
of $B \ra X_s \gamma$ Decay}{\bf\large$^*$}
\vspace{0.3in} \\
\setlength {\baselineskip}{0.2in}

{\large  A. J. Buras,}$^{1,2)}$
{\large \ M. Misiak,}$^{1)}{}^{\dagger}$
{\large \ M. M\"unz}$^{1)}$
{\large \ and \ S. Pokorski}$^{2)}{}^{\dagger}$
\vspace{0.3in}

$^{1)}${\it Physik-Department\\
Technische Universit\"at M\"unchen\\
D-85747 Garching, Germany\\}
\vspace{0.3in}

$^{2)}${\it Max-Planck-Institut f\"ur Physik\\
  Werner-Heisenberg-Institut\\
F\"ohringer Ring 6\\ D-80805 M\"unchen,
  Germany\\}

\vspace{1in}
{\bf ABSTRACT \\}
\vspace{0.1in}
  \end{center}

We analyze uncertainties in the theoretical prediction for the
inclusive branching ratio $BR[B \ra X_s \gamma]$. We find that
the dominant uncertainty in the leading order expression comes from
its $\mu$-dependence. We discuss a next-to-leading order calculation
of $B \ra X_s \gamma$ in general terms and check to what
extent the $\mu$-dependence can be reduced in such a calculation.  We
present constraints on the Standard and Two-Higgs-Doublet Model
parameters coming from the measurement of $b \ra s
\gamma$ decay. The current theoretical uncertainties do not
allow one to definitively restrict the Standard Model parameters much
beyond the limits coming from other experiments. The bounds on the
Two-Higgs-Doublet Model remain very strong, though significantly
weaker than the ones present in the recent literature. In the
Two-Higgs-Doublet Model case, the $b \ra s \gamma$, $Z \ra b \bar{b}$
and $b \ra c \tau \bar{\nu}_{\tau}$ processes are enough to give the
most restrictive bounds in the $M_{H^{\pm}}-tan \beta$ plane.

\vspace{0.5in}
\noindent \underline{\hspace{2in}}\\
$^{\dagger}$ {\footnotesize On leave of absence from Institute of
Theoretical Physics, Warsaw University.}\\
$^*$ {\footnotesize Supported in part by the German Bundesministerium
f\"ur Forschung und Technologie under contract 06 TM 732 and by the
Polish Commitee for Scientific Research.}
\end{titlepage}

\noindent {\bf 1. Introduction}

\begin{figure}[htb]
\rotate[r]{
\epsfysize = 0.9in
\epsffile{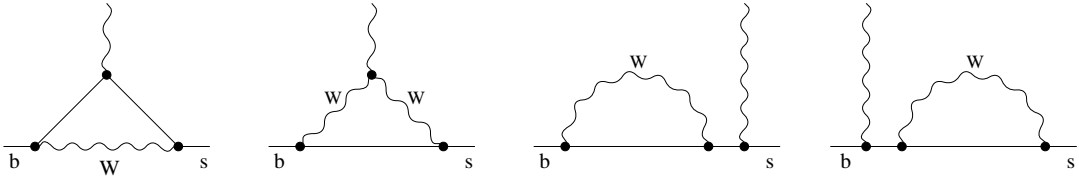}
}
\vspace{-5.3in}
\caption{Leading order one loop diagrams generating the \bsg
transition in the SM.}
\label{SMdiag}
\end{figure}

\indent	The \bsg decay is known to be extremely sensitive to the
structure of fundamental interactions at the electroweak scale. As any
Flavour Changing Neutral Current (FCNC) process, it does not arise at
the tree level in the Standard Model (SM). The one-loop W-exchange
diagrams that generate this decay at the lowest order in the SM
(fig. \ref{SMdiag}) are small enough to be comparable to possible
nonstandard contributions (charged scalar exchanges, SUSY one loop
diagrams, $W_R$ exchanges in the L-R symmetric models, etc.).\\

        Among all FCNC processes, the \bsg decay is particularly
interesting because its rate is of order $G_F^2 \alpha_{\scs QED}$,
while most of the other FCNC processes involving leptons or photons
are of order $G_F^2 \alpha_{\scs QED}^2$.  The long-range strong
interactions are expected to play a minor role in the inclusive
\Bsg decay.\footnote{$X_s$ can be identified by requiring that
$E_{\gamma} > (m_B^2 - m_D^2)/2m_B$} This is because the mass of the
b-quark is much larger than the QCD scale $\Lambda$. Moreover, the
only relevant intermediate hadronic states $\psi X_s$ are expected to
give very small contributions, as long as we assume no interference
between short- and long-distance terms in the inclusive rate
\cite{psi}. Therefore, it has become quite common to use the following
approximate equality to estimate the
\Bsg rate

\be
\f{\Gamma[B \ra X_s \gamma]}
     {\Gamma[B \ra X_c e \bar{\nu}_e]}
 \simeq                                                     \label{ratios}
\f{\Gamma[b \ra s \gamma]}
     {\Gamma[b \ra c e \bar{\nu}_e]} \equiv R(m_t,\al,\xi)
\ee

\noindent where the quantities on the r.h.s are calculated in the
spectator model corrected for short-distance QCD effects. The
normalization to the semileptonic rate is usually introduced in order
to cancel the uncertainties due to the Cabibbo-Kobayashi-Maskawa (CKM)
matrix elements and factors of $m_b^5$ in the r.h.s. of
eq. (\ref{ratios}).

	As indicated above, this ratio depends only on $m_t$ and $\al$
in the Standard Model. In the extensions of the SM, additional
parameters are present. They have been commonly denoted by $\xi$. The
main point to be stressed here is that R is a calculable function of
its parameters in the framework of a renormalization group improved
perturbation theory. Consequently, the decay in question is
particularly suited for the tests of the SM and its extensions.

 One of the main difficulties in analyzing the inclusive
\Bsg decay is calculating the short-distance QCD effects due to hard
gluon exchanges between the quark lines of the leading one-loop
electroweak diagrams. These effects are known
\cite{BertDesh}--\cite{Ciu} to enhance the \bsg rate in the SM by 2--5
times, depending on the top quark mass. So the \bsg decay appears to
be the only known process in the SM that is dominated by two-loop
contributions.\footnote{apart from other $q \ra q' \gamma$
transitions which are very hard to measure}

        With the above conclusion in mind, one can realize how
astonishing it is that the \bsg decay is already measured! The recent
CLEO report \cite{CLEO} gives the following branching ratio for the
exclusive $B \ra K^* \gamma$ decay

\be
BR[B \ra K^* \gamma] = (4.5 \pm 1.5 \pm 0.9) \times 10^{-5}.
\label{excl}
\ee

\noindent Therefore, we can say that the inclusive \Bsg branching ratio is
measured to be larger than the one in eq. (\ref{excl}), and smaller
\cite{CLEO} than

\be
BR[B \ra X_s \gamma] < 5.4 \times 10^{-4} \hspace{.5in}
(at\;\;95\%\;\; C.L.).                          \label{incl}
\ee

These experimental findings are in the ball park of the SM
expectations based on the leading logarithmic approximation. It is
then not surprising that after announcement of the results
(\ref{excl}) and (\ref{incl}) several analyses appeared in the
literature with the purpose to restrict the parameters of the SM or
put bounds on its extensions.

	In view of these new developments, it is the purpose of the
present paper to reanalyze the theoretical uncertainties in the
calculation of the ratio (\ref{ratios}). We find, in agreement with
ref. \cite{AG1}, that the dominant uncertainty in the existing leading
logarithmic calculations is due to choice of the renormalization scale
$\mu$. Such an uncertainty, inherent in any finite order of
perturbation theory, has been recently analyzed in several FCNC
processes, such as $B^0$-$\bar{B}^0$ and $K^0$-$\bar{K}^0$ mixing or
rare K- and B-decays \cite{A1,A2,A3}. It has been demonstrated in
these papers, that the inclusion of next-to-leading order corrections
reduces considerably the $\mu$-dependence of the relevant amplitudes.

	Since \Bsg is dominated by QCD effects, it is not surprising
that the scale-uncertainty in the leading order calculation of this
decay rate is particularly large --- it amounts to around $\pm 25\%$. It
follows that the restrictions on the SM or its extensions which can be
obtained with help of the experimental findings (\ref{excl}) and
(\ref{incl}) and the leading order approximation are substantially
weaker than found by other authors.

	Unfortunately, it will take some time before the
$\mu$-dependences present in \Bsg can be reduced in the same manner as
it was done for the other decays \cite{A1,A2,A3}. As we will describe
in the following, a full next-to-leading calculation of \Bsg would
require calculation of three-loop mixings between certain effective
operators.  Before one undertakes such an effort, it is instructive to
make a formal analysis of the considered decay at the next-to-leading
level and to check to what extent the $\mu$-dependence can be reduced
once all the necessary calculations have been performed.

	Our paper is organized as follows: In section 2 we summarize
the results of the leading logarithmic calculations. Subsequently, we
discuss various uncertainties present in the existing formulae. In
section 3 we analyze the restrictions from eq. (\ref{incl}) on the
parameters of the Standard and Two-Higgs-Doublet Models, taking into
account the uncertainties discussed in the previous section.  In
section 4 we generalize the renormalization group analysis beyond the
leading logarithmic approximation, we list the calculations which have
to be done (or have been already done), and demonstrate how the
$\mu$-dependence will be reduced. Section 5 gives a brief summary of
the paper.\\
\ \\
\noindent {\bf 2. Present status and theoretical uncertainties}\\
\ \\
\indent Let us start by summarizing the results of the leading
logarithmic calculations \cite{Grin}--\cite{Ciu}. In a compact form,
they can be written as follows:

\be                   \label{main}
R =
\f{\Gamma[b \ra s \gamma]}{\Gamma[b \ra c e \bar{\nu}_e]}
 =  \f{|V_{ts}^* V_{tb}|^2}{|V_{cb}|^2}
\f{6 \alpha_{\scs QED}}{\pi g(z)} |C^{(0)eff}_7(\mu)|^2
\ee

\noindent where

\be        \label{c7eff}
C^{(0)eff}_7(\mu) = \eta^{\f{16}{23}} C^{(0)}_7(M_W) +
\f{8}{3} \left( \eta^{\f{14}{23}} - \eta^{\f{16}{23}}
\right) C^{(0)}_8(M_W)  + C^{(0)}_2(M_W) \sum_{i=1}^8 h_i \eta^{a_i}
\ee

\noindent with $z = \f{m_c}{m_b}$, $\eta = \al(M_W)/\al(\mu)$, and

\be
g(z) = 1 - 8z^2 + 8z^6 - z^8 - 24z^4 ln \; z.           \label{g}
\ee

The function $g(z)$ is the phase space factor in the semileptonic
b-decay.\footnote{Note, that at this stage we do not include the
${\cal O}(\al)$ corrections to $\Gamma(b \ra c e \bar{\nu})$ since they are
part of the next-to-leading effects discussed in section 4.} The
numbers $h_i$ and $a_i$ are given and discussed in the appendix. Here
we only mention that the sum of all the $h_i$ vanishes. Finally, the
Standard Model values of the coefficients $C^{(0)}_i(M_W)$ are
\cite{InamiLim}

\be
C^{(0)}_2(M_W) = 1                               \label{c2}
\ee
\be
C^{(0)}_7(M_W) = \f{3 x^3-2 x^2}{4(x-1)^4}ln x + \f{-8 x^3 - 5 x^2 + 7
x}{24(x-1)^3}                               \label{c7}
\ee
\be
C^{(0)}_8(M_W) = \f{-3 x^2}{4(x-1)^4}ln x + \f{-x^3 + 5 x^2 + 2
x}{8(x-1)^3}                                \label{c8}
\ee

\noindent where $x = \f{m_t^2}{M_W^2}$. Some details of the derivation
of eqs. (\ref{main})--(\ref{c8}), as well as the reason for introducing
the symbol $C^{(0)eff}_7(\mu)$ will be given in section 4.

	The renormalization scale $\mu$ present in eqs. (\ref{main})
and (\ref{c7eff}) has to be of order $m_b$, but need not to be exactly
equal to $m_b$. The QCD coupling constant $\al(\mu)$ at any
renormalization scale $\mu$ can be expressed (as it has recently
become popular) in terms of its value at $\mu = M_Z$. The
leading logarithmic expression is just $\al(\mu) = \al(M_Z)/v(\mu)$,
where

\be \label{v(mu)}
v(\mu) = 1 - \beta_0 \f{\al(M_Z)}{2 \pi} ln \left( \f{M_Z}{\mu}
\right) \label{v}
\ee

\noindent and $\beta_0 = 11 - \f{2}{3}f$. In our case, the number of active
flavors $f$ equals 5.

        Before starting the discussion of uncertainties, let us
illustrate the relative numerical importance of the three terms in the
expression (\ref{c7eff}) for $C^{(0)eff}_7$. For instance, for $m_t
= 130 GeV$, $\mu = 5 GeV$ and $\al(M_Z)=0.12$ one obtains

\begin{eqnarray}
C^{(0)eff}_7(\mu)=0.689\;C^{(0)}_7(M_W)\;\;+\;\;0.087\;C^{(0)}_8(M_W)
\;\;-\;\;0.161\;C^{(0)}_2(M_W) =\nonumber\\=
0.689\;(-0.161)\;\;+\;\;0.087\;(-0.086)\;\;-0.161\;\;= -0.280.
\end{eqnarray}

\noindent In the absence of QCD we would have $C^{(0)eff}_7(\mu) =
C^{(0)}_7(M_W)$ (in that case one has $\eta = 1$). Therefore, the
dominant term in the above expression (the one proportional to
$C^{(0)}_2(M_W)$) is the additive QCD correction that causes the
enormous QCD-enhancement of the \bsg rate.  It originates solely from
the two-loop diagrams. On the other hand, the multiplicative QCD
correction (the factor 0.689 above) tends to suppress the rate, but
fails in the competition with the additive contributions.

         The equations (\ref{ratios}) and (\ref{main})--(\ref{g})
summarize the complete leading logarithmic (LL) expression for the
\Bsg rate in the SM.  Their important property is that they are
exactly the same in many interesting extensions of the SM, such as the
Two-Higgs-Doublet Model (2HDM) or the Minimal Supersymmetric Standard
Model (MSSM).\footnote{However, in a general $SU(2)_L \times SU(2)_R
\times U(1)$ model they get modified, because some new ``effective
operators'' enter - see ref. \cite{LR}.} The only quantities that
change are the coefficients $C^{(0)}_2(M_W)$, $C^{(0)}_7(M_W)$ and
$C^{(0)}_8(M_W)$ (see refs. \cite{Grin,Borzum} for details).

\ \\
	Let us now have a critical look at eqs. (\ref{ratios}) and
(\ref{main})--(\ref{v}), and list the theoretical and experimental
uncertainties present in the prediction for BR[\Bsg] that can be made
with help of these equations.

\ \\
(i) First of all, eq. (\ref{ratios}) is based on the spectator
model. For a long time there has been a question of how good this
model is and what is its relation to QCD. Fortunately, during the past
two years, this relation has been better understood. The spectator
model has been shown to correspond to the leading order approximation
in $1/m_b$ expansion. The next corrections appear at the ${\cal O}(1/m_b^2)$
level. The latter terms have been studied by several authors
\cite{spect} with the result that they affect BR[\Bsg] and BR[$B \ra
X_c e \bar{\nu}_e$] by only a few percent, as long as one uses
$m_b=m^{pole}_b$ in the leading term. In view of much larger
uncertainties discussed below, the exact size of the ${\cal
O}(1/m_b^2)$ corrections is not essential. The effects due to the
difference between $m^{pole}_b$ and $m_b(\mu=m_b)$ are next-to-leading
QCD effects which we discuss in point (v) below, and in section 4. In
our phenomenological analysis presented in the next section, we will
assume the error due to the use of spectator model in
\Bsg to amount to $\pm 10\%$. This error is also meant to include
interference between the short-distance contributions and
contributions from the $\psi X_s$ intermediate states \cite{psi}. It
could be significantly larger, had interference terms in all the
exclusive channels the same signs, but we can see no reason for such a
correlation.

\ \\
(ii) As already mentioned, the normalization of the \Bsg rate to the
semileptonic rate in eq. (\ref{ratios}) has been introduced in order
to cancel the factors of $m_b^5$ on the r.h.s. of this equation. It
also drastically reduces uncertainties due to the CKM matrix elements
(see point (iii) below). The price to pay is an additional uncertainty
coming from the ratio $z = \f{m_c}{m_b}$ in the phase-space factor
for the semileptonic decay.  However, this ratio is experimentally
known better than the actual masses of both quarks. And this is the
only combination in which these two masses enter our expressions
(\ref{ratios}) and (\ref{main})--(\ref{v}).

	The phase-space factor $g(z)$ changes rather fast with $z$
around $z=\f{1}{3}$. In the next section we will use $z = 0.316 \pm
0.013$ \cite{Ruckl}. The resulting error in the ratio R is then around
6\%. The actual accuracy of determining $z$ can of course be subject
to discussion. However, even if one assumes that the error due to $z$
is twice larger than 6\%, it still remains significantly smaller than
the ones we discuss below.

\ \\
(iii) The error due to the ratio of the CKM parameters in
eq. (\ref{main}) is small. Assuming unitarity of the $3 \times 3$ CKM
matrix and imposing the constraint from the CP-violating parameter
$\epsilon_{\scs K}$ we find
\be
\f{|V_{ts}^* V_{tb}|^2}{|V_{cb}|^2} = 0.95 \pm 0.04     \label{KM}
\ee

\begin{figure}[htb]
\centerline{
\rotate[r]{
\epsfysize = 4in
\epsffile{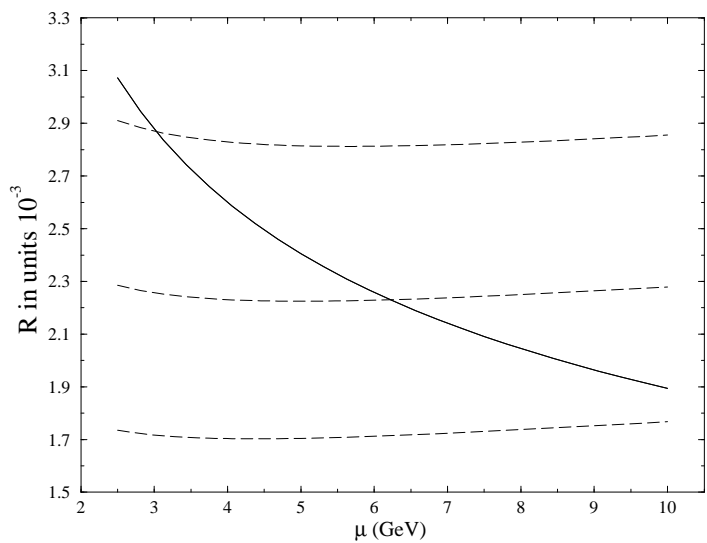}
}}
\caption{$\mu$-dependence of the theoretical prediction for the ratio
R, for $m_t$=150 GeV and $\al(M_Z)=0.12$. The solid line corresponds
to the leading order prediction. The dashed lines describe possible
next-to-leading results (see section 4).}
\label{mudep}
\end{figure}

The quoted error corresponds to $0.036 \leq |V_{cb}| \leq 0.047$
\cite{Lueth}, $|V_{ub}/V_{cb}| = 0.08 \pm 0.02$ and $B_K = 0.7 \pm
0.2$.  Eq. (\ref{KM}) has been obtained for $m_t$ set to 200 GeV. For
smaller values of $m_t$ the central value is practically the same, and
the error decreases down to 0.03 for $m_t=100\;GeV$.

\ \\
(iv) There exists an uncertainty due to the determination of
$\al$. This uncertainty is not small because of the importance of QCD
corrections in the considered decay. Now all the extractions of $\al$
which give conservatively the range $0.11 \leq \al(M_Z) \leq 0.13$
\cite{alpha} in the $\overline{MS}$ scheme, include next-to-leading
order corrections both in the expressions for $\al$ and in the
relevant processes used for $\al$ determination. Unfortunately, the
next-to-leading order corrections to
\Bsg are unknown at present. So for this decay it is not possible to
consistently use $\al(M_Z)$, say in $\overline{MS}$ scheme. For this
reason, the usage of two-loop expressions for $\eta$ in
eq. (\ref{c7eff}) can certainly be questioned. However, if one takes
the same initial condition $\al(M_Z) = \alpha_{\overline{MS}}(M_Z)$
for the evolution of $\al(\mu)$, then the difference between $R$'s in
eq. (\ref{main}) resulting from using the two-loop and the one-loop
expressions for $\al$ amounts only to 4--9\%. On the other hand, the
difference between the ratios R of eq.  (\ref{main}) obtained with
help of $\alpha_{\overline{MS}}(M_Z)=0.11$ and $0.13$, respectively,
is between 15 and 22\%. So the inaccuracy in determination of
$\al(M_Z)$ is more important here than the NLL effects in the
evolution of $\al(\mu)$. In the next section we will use just the
leading logarithmic expressions for $\al$ and include only errors due
to varying $\al(M_Z)$ between 0.11 and 0.13.

\ \\
(v) The dominant uncertainty in eq. (\ref{ratios}) comes from the
unknown next-to-leading order contributions. This uncertainty is best
signaled by the strong $\mu$-dependence of the leading order
expression (\ref{main}), which is shown by the solid line in
fig. \ref{mudep}, for the case $m_t=150\;GeV$.

	One can see that when $\mu$ is varied by a factor of 2 in both
directions around $m_b \simeq 5\;GeV$, the ratio (\ref{main}) changes
by around $\pm 25\%$, i.e. the ratios R obtained for $\mu=2.5\;GeV$
and $\mu=10\;GeV$ differ by a factor of 1.6. This is in a rough
agreement with what has been found in ref. \cite{AG1}. In section 4 we
will argue why varying $\mu$ by a factor of 2 in both directions
properly estimates the size of unknown next-to-leading QCD
corrections.

	The dashed lines in fig. \ref{mudep} show the expected
$\mu$-dependence of the ratio (\ref{main}) once a complete
next-to-leading calculation is performed. The $\mu$-dependence is then
much weaker, but until one performs the calculation explicitly one
cannot say which of the dashed curves is the proper one. The way the
dashed lines are obtained is described in section 4.

	For the purpose of the next section, we will estimate the size
of the unknown next-to-leading terms by varying $\mu$ by a factor of 2
in both directions around $m_b \simeq 5\;GeV$ in the existing leading
logarithmic expressions (\ref{main})--(\ref{c7eff}).

	There is one more comment we should make at this point: The
scale used in the numerator of the ratio $\eta=\al(M_W)/\al(\mu)$ (the
``matching'' scale) does not have to be exactly equal $M_W$ but rather
can also be varied by a factor of 2 around $M_W$. The resulting
changes in the ratio R are then roughly twice smaller than the ones we
have found by varying the low-energy scale around $m_b$. However, this
uncertainty is also due to next-to-leading effects. So we will treat
it as already taken into account in the estimate we have made by
varying the low-energy scale, i.e. we will not vary the matching scale
in our analysis described in the next section.

\ \\
(vi) Finally, we have to mention that there exists a 4.6\% error in
determining $BR[B \ra X_s \gamma]$ from eq. (\ref{ratios}), which is
due to the error in the experimental measurement of $BR[B\ra X_c e
\bar{\nu}_e] = (10.7 \pm 0.5)\%$ \cite{PData}.

\ \\
We treat the $m_t$-dependence in eqs. (\ref{c7})--(\ref{c8}) as an
``interesting uncertainty'', i.e. all our predictions will be given as
functions of $m_t$. We will never add this uncertainty to the other
errors.

        The above discussion of uncertainties is to a large extent
model-independent. None of the errors listed in points (i)--(vi) was
due to the coefficients $C^{(0)}_2(M_W)$, $C^{(0)}_7(M_W)$ and
$C^{(0)}_8(M_W)$ (except for NLL contributions to them).\footnote{The
effect of the QCD evolution between the $m_t$ and $M_W$ scales can be
treated as NLL contribution as long as $ln(m_t/M_W) \leq 1$, i.e. when
$m_t \leq 220 GeV$. This contribution has been already calculated in
ref.\cite{topW} and found to give around 10\% effect.}  Therefore, in
all extensions of the SM where only $C^{(0)}_i(M_W)$ get modified
(like in the 2HDM or the MSSM), the above estimates remain valid. Of
course, in extensions of the SM, some additional uncertainties in
$C^{(0)}_i(M_W)$ may occur. Apart from the dependence of
$C^{(0)}_i(M_W)$ on the parameters of these models (like
e.g. $tan\beta$ or charged scalar mass in the 2HDM), which is a
welcome feature, we can also encounter additional uncertainties due to
the SM parameters: For instance, $C^{(0)}_i(M_W)$ in the MSSM contain
ratios of the ordinary and SUSY-sector mixing angles (see
ref. \cite{Borzum}).\\
\ \\
\noindent {\bf 3. Phenomenological consequences for the Standard and
Two-Higgs-Doublet Models}\ \\
\ \\
\indent In the present section we describe in what manner the
inclusion of theoretical uncertainties affects the limits on the
Standard and Two-Higgs-Doublet Model parameters that can be obtained
from \Bsg measurement.

\begin{figure}[htb]
\centerline{
\rotate[r]{
\epsfysize = 4in
\epsffile{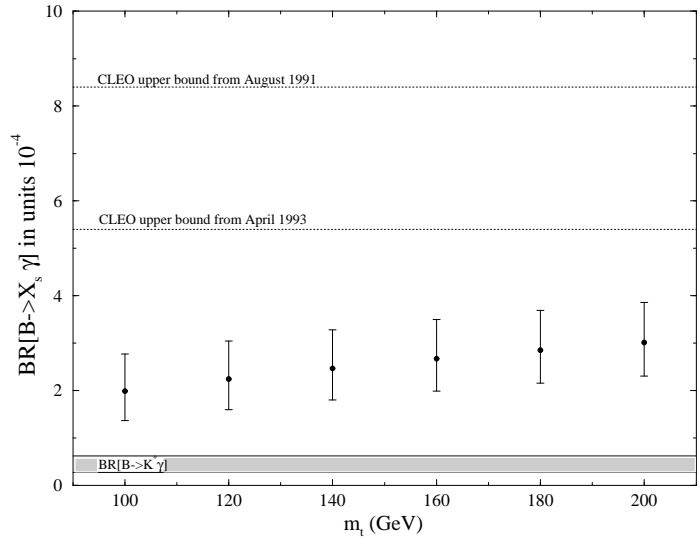}
}}
\caption{Predictions for $B\rightarrow X_s\gamma$ in the SM, as a
function of the top quark-mass, with the theoretical uncertainties
taken into account (see the text).}
\label{barsSM}
\end{figure}

	Fig. \ref{barsSM} presents the SM prediction for the inclusive
\Bsg branching ratio. In order to obtain the error bars presented in
this figure, we have added in quadratures the errors (i)--(vi) listed in
the previous section. Adding in quadratures does not mean, of course,
that we can attribute any statistical meaning to these
errors. However, it is the best one can do at present in order to
estimate the total uncertainty in the theoretical prediction for
BR[\Bsg]. By adding in quadratures we want to take into account that
there is no correlation between the errors (i)--(vi).

        Comparing the theoretical SM prediction with the experimental
results in fig. \ref{barsSM}, one can conclude that it is rather
improbable that the future measurements will contradict this
prediction, in view of its large uncertainty. Comparison with
experiment can become sharper only by making the theoretical
prediction more precise --- first of all by calculating the complete
NLL short-distance effects.

        One can also ask a related but somewhat different question of
whether, assuming the present theoretical uncertainty, a future
precise measurement of the inclusive \Bsg rate may tell us something
about the SM parameters.

\begin{figure}[htb]
\centerline{
\rotate[r]{
\epsfysize = 4in
\epsffile{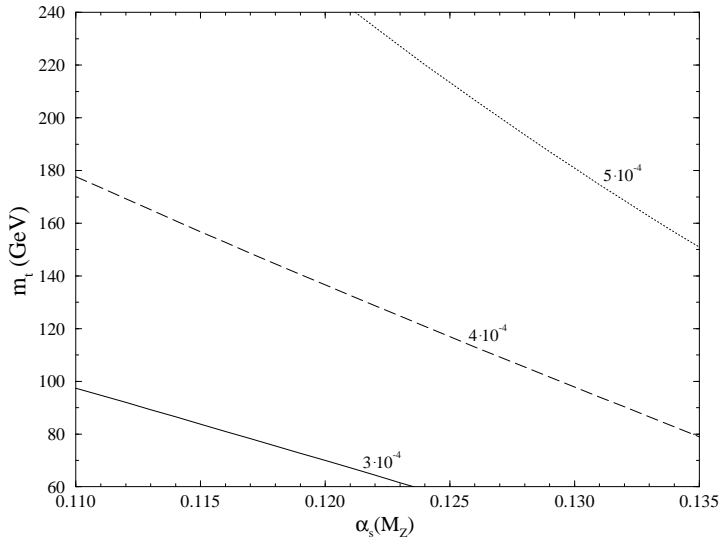}
}}
\caption{Lower limits for $m_t$ in the SM as a function of
$\al(M_Z)$ (see the text).}
\label{mtalfalow}
\end{figure}

        Fig. \ref{mtalfalow} presents the lower limits on the top-quark
mass following from the assumption that the future experimental result
will satisfy the inequality

\be  \label{2s_low}
BR^{theor}[B \ra X_s \gamma]\;\;\;\; + \;\;\;\;2 \times
\left( \begin{array}{c}{\rm theoretical}\\
                       {\rm error\;\;\;bar} \end{array} \right)
\;\;\;\; < \;\;\;\;
\left( \begin{array}{c}{\rm measured}\\
                       {\rm decay\;\;\; rate} \end{array} \right)
\ee

\noindent for a certain value of $m_t$. Such $m_t$ is then treated as
disallowed. The three curves correspond to different possible results
of the measurement of BR[\Bsg] (3, 4, and 5 times $10^{-4}$).

        If we trust our estimation of the theoretical uncertainty, the
conclusion is that we may get a useful lower bound on the top quark
mass provided the branching ratio is close to the present experimental
upper bound. It has to be stressed, however, that the limits in
fig. \ref{mtalfalow} are quite sensitive to the treatment of the
errors. Even if one only symmetrized our errors in fig. \ref{barsSM},
these limits would move up by around 10\%.

        One could also ask what upper limits on the top quark mass (as
functions of $\al(M_Z)$) can be obtained by assuming BR[\Bsg] is
measured below the theoretical prediction, and the inequality

\be  \label{2s_up}
BR^{theor}[B \ra X_s \gamma]\;\;\;\; - \;\;\;\;2 \times
\left( \begin{array}{c}{\rm theoretical}\\
                       {\rm error\;\;\;bar} \end{array} \right)
\;\;\;\; > \;\;\;\;
\left( \begin{array}{c}{\rm experimental}\\
                       {\rm upper\;\;\;bound} \end{array} \right)
\ee

is satisfied. We find that upper bounds for $m_t$ that lie below 200
GeV can be obtained only if one assumes that the inclusive rate will
be measured smaller than around $1.7 \times 10^{-4}$. This is not very
encouraging, in view of the present measurement quoted in
eq. (\ref{excl}), and the expectation that the $B \ra K^* \gamma$ rate
forms 10--20\% of the inclusive rate \cite{AG2,bkg}, depending on the
model used for making such an estimate.

	Let us now discuss the limits on the Two-Higgs-Doublet Model.

 	The relevant part of the 2HDM lagrangian \cite{2HDM} which is
responsible for the nonstandard contributions to \bsg has the
following form:

\be
L_H = \sqrt{\f{4 G_F}{\sqrt{2}}} H^+ \left[ X \bar{U}_L V_{CKM} M_D D_R
        +   Y \bar{U}_R M_U V_{CKM} D_L \right] + h.c..
\ee

\noindent Here $H^+$ is the charged scalar field, U and D represent the column
vectors of the up- and down-quarks, respectively, $V_{CKM}$ is the CKM
matrix, and $M_U$ ($M_D$) is the diagonal mass matrix for the up-
(down-) quarks. The quantities X and Y are given by $tan\beta =
v_2/v_1$, i.e. the ratio of the vacuum expectation values of the two
scalar doublets. In the most popular version of the 2HDM, usually
called ``Model II'', the masses of the down-quarks (up-quarks) are
proportional to $v_1$ $(v_2)$.\footnote{The notation in
ref.\cite{Grin} is opposite.} Then one has

\be
X = \f{1}{Y} = tan\beta = \f{v_2}{v_1}.
\ee
\ \\
\begin{figure}[htb]
\rotate[r]{
\epsfysize = 0.9in
\epsffile{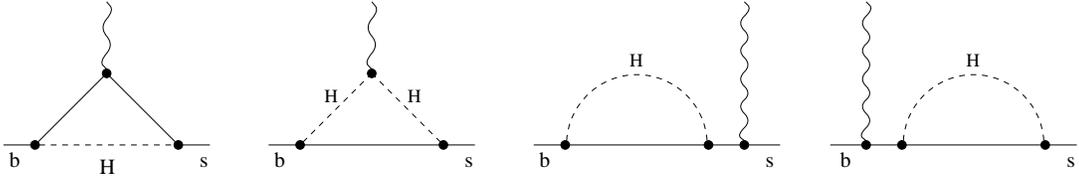}
}
\vspace{-5.3in}
\caption{Charged scalar exchanges contributing to \bsg in the 2HDM.}
\label{2HDMdiag}
\end{figure}

        The charged scalar exchanges shown in fig. \ref{2HDMdiag}
create extra contributions to the coefficients $C^{(0)}_7(M_W)$ and
$C^{(0)}_8(M_W)$ (see ref. \cite{Grin}):

\begin{eqnarray}
\Delta C^{(0)}_7(M_W)   =   \f{ 3y^2 - 2y}          { 6(y-1)^3}ln y  +
                            \f{-5y^2 + 3y}          {12(y-1)^2}      +
\f{1}{tan^2\beta} \left[    \f{3 y^3 - 2y^2}        {12(y-1)^4}lny   +
                            \f{-8 y^3 - 5 y^2 + 7y} {72(y-1)^3} \right]
\nonumber\\ \nonumber \\ \label{extra} \\ \nonumber \\
\Delta C^{(0)}_8(M_W)   =   \f{-y}                  {2(y-1)^3}ln y   +
                            \f{-y^2 + 3y}           {4(y-1)^2}       +
\f{1}{tan^2\beta} \left[    \f{-y^2}                {4(y-1)^4}ln y   +
                            \f{-y^3 + 5 y^2 + 2y}   {24(y-1)^3}    \right]
\nonumber
\end{eqnarray}

\noindent where $y=m_t^2/M_{H^{\pm}}^2$.

\begin{figure}[htb]
\centerline{
\rotate[r]{
\epsfysize = 4in
\epsffile{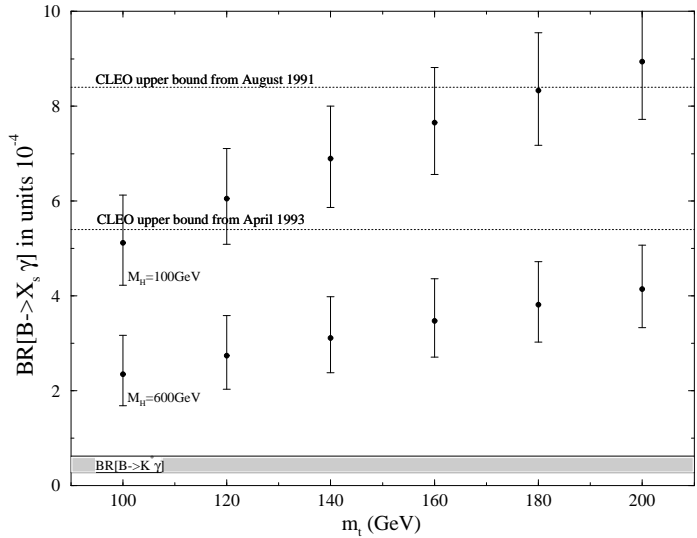}
}}
\caption{Predictions for $B\rightarrow X_s\gamma$ in the 2HDM, as a
function of the top-quark mass, with the theoretical uncertainties
taken into account (see the text).}
\label{bars2HDM}
\end{figure}

        As we have already said, equations (\ref{main})--(\ref{g})
remain unaltered. So from eq. (\ref{extra}) we can immediately obtain
the 2HDM predictions for the \Bsg branching ratio. They are presented
in fig. \ref{bars2HDM}, for the case $tan\beta=2$ which is large enough
to make the terms proportional to $1/tan^2\beta$ in eq. (\ref{extra})
practically irrelevant. For smaller $tan\beta$, the results grow
rapidly as $tan\beta$ decreases.

        For most of the top-quark masses, the results for $M_{H^{\pm}}
= 100 GeV$ in fig. \ref{bars2HDM} are in contradiction with present
measurements, even if the theoretical uncertainties are taken into
account. On the other hand, comparing figs. \ref{barsSM} and
\ref{bars2HDM} we can conclude that in the $M_{H^{\pm}} = 600 GeV$
case we are not able to distinguish between the SM and the 2HDM,
independently of the accuracy of future measurements (but assuming the
present accuracy of the theoretical predictions).

\begin{figure}[htb]
\centerline{
\rotate[r]{
\epsfysize = 4in
\epsffile{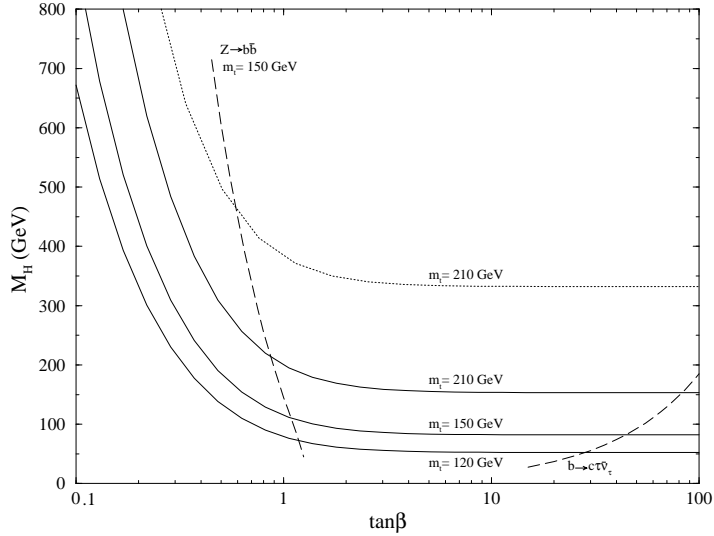}
}}
\caption{Lower limits for the charged Higgs boson mass as a function of
$\tan\beta$ in the 2HDM (see the text).}
\label{mhtanb}
\end{figure}

        Fig. \ref{mhtanb} presents the limits in the
$tan\beta$-$M_{H^{\pm}}$ plane, obtained by assuming that the excluded
values of $tan\beta$ and $M_{H^{\pm}}$ are those for which the
inequality (\ref{2s_up}) holds. The presented limits are found with
help of the experimental input from eq. (\ref{incl}). The three solid
curves represent lower bounds on $M_{H^{\pm}}$ obtained from \bsg for
three different values of the top quark mass: 120, 150 and 210
GeV. The dashed curves show bounds coming from the $Z \ra b \bar{b}$
and $b \ra c \tau \bar{\nu}_{\tau}$ decays.  The shown $Z \ra b
\bar{b}$ bound is based on the results of ref. \cite{Hollik} and
corresponds to $m_t=150\;GeV$. The $b \ra c \tau \bar{\nu}_{\tau}$
bound is given by the simple formula
\cite{Isidori}

\be
\tan\beta < 0.54 \f{M_H}{1 GeV}.
\ee

\noindent The $b \ra c \tau \bar{\nu}_{\tau}$ and $Z \ra b \bar{b}$
bounds are complementary to the \bsg bound. All other bounds in the
$tan\beta$-$M_{H^{\pm}}$ plane (e.g. the ones coming from
$B^0$-$\bar{B}^0$ mixing and other FCNC processes considered in
refs. \cite{AB2HDMandGG}) are weaker than the combination of the three
above-mentioned ones.

	Our \bsg bounds are much weaker than the ones found in
refs. \cite{Hew,BPh}, because we have taken into account the
uncertainties in the theoretical prediction. If we did not take them
into account, our bound for $m_t$ = 210 GeV would lie as high as the
dotted curve in fig. \ref{mhtanb}.

\begin{figure}[htb]
\centerline{
\rotate[r]{
\epsfysize = 4in
\epsffile{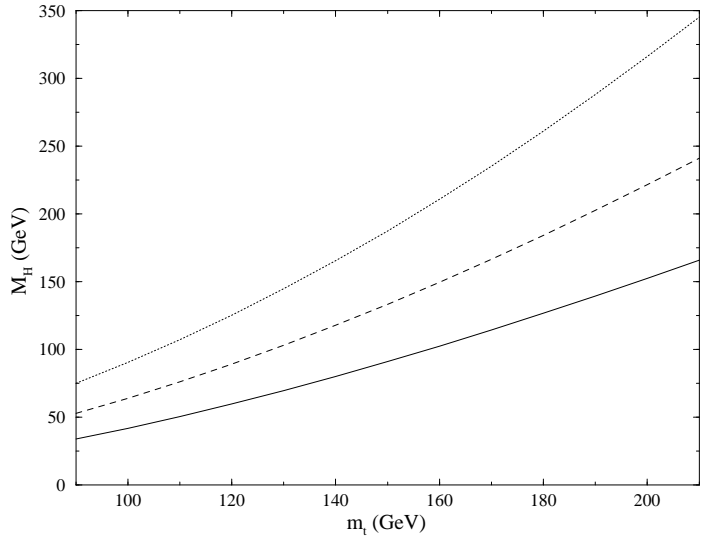}
}}
\caption{The \bsg lower limits for the charged Higgs boson mass as functions
of $m_t$ for $\tan\beta = 2$. See the text for the description of the
curves.}
\label{mhmt}
\end{figure}

	As one can see, the \bsg bounds in fig. \ref{mhtanb} are
practically independent of $tan\beta$ for its values larger than
2. From the theoretical standpoint, large values of $tan\beta$ are
interesting  since $tan\beta \simeq m_t/m_b$ would naturally explain
the large splitting between these two masses without necessity of
introducing order-of-magnitude different Yukawa couplings \cite{tanb}.
For these $tan\beta$'s it is convenient to plot the
$tan\beta$-independent bounds in the $m_t$-$M_{H^{\pm}}$ plane. They
are presented in fig. \ref{mhmt}.

The solid line in this figure shows the lower bound for $M_{H^{\pm}}$
obtained using the requirement described in eq. (\ref{2s_up}). If we
require the theoretical prediction for BR[\Bsg] be only one, not two
error bars away from the experimental upper limit, then our lower
bound on $M_{H^{\pm}}$ is described by the dashed curve in
fig. \ref{mhmt}. The dotted line in this figure corresponds to assuming
that the theoretical prediction is given by its central value, with no
uncertainties whatsoever. We present the three curves in order to
demonstrate the relevance of taking the theoretical uncertainties into
account.

	Our limits are much weaker than found in ref. \cite{Hew}, and
they do not exclude the $t \ra H^+ b$ decay of the top-quark.
However, they still have to be considered very strong. For the most
interesting values of $tan\beta$ they are still more restrictive than
the bounds from the $Z \ra b\bar{b}$ decay \cite{Hollik,Park}, as well
as the bounds from other FCNC processes \cite{AB2HDMandGG}. The
expected new data on \Bsg in the nearest future will further improve
the bounds on the 2HDM, but the possible improvement is limited by the
theoretical uncertainties. Comparison between figs. \ref{barsSM} and
\ref{bars2HDM} shows that for $M_{H^{\pm}}$ larger than around 600
GeV, the present theoretical inaccuracy does not allow to distinguish
the 2HDM and SM predictions.\\
\ \\
\noindent {\bf 4. The structure of \Bsg beyond leading logarithms}\\
\ \\
\indent In this section we describe a complete next-to-leading
calculation of \Bsg in general terms, and check to what extent the
theoretical uncertainties due to the $\mu$-dependence can be reduced
after performing such a calculation.

	The QCD corrections to the \Bsg decay contain large logarithms
$ln(M^2_W/m^2_b)$ which have to be resummed with help of the
Renormalization Group Equations (RGE). In order to do this, one
introduces an effective hamiltonian built out of operators of
dimension higher than 4.
\be \label{ham}
H_{eff} =
-V_{tb} V^*_{ts} \f{G_F}{\sqrt{2}} \sum_{i=1}^{8} Q_i(\mu) C_i(\mu)
\equiv
-V_{tb} V^*_{ts} \f{G_F}{\sqrt{2}} \vec{Q}^T(\mu) \vec{C}(\mu)
\ee
\noindent where $V_{ij}$ are the elements of the CKM matrix, $Q_i(\mu)$ are the
relevant operators and $C_i(\mu)$ are the corresponding Wilson
coefficients. The complete set of operators necessary in the \bsg case
is the following \cite{Grin}:

\be \label{ope}
\begin{array}{cc}
Q_1 = &  (\bar{s}_{\alpha}  c^{\beta })_{V-A}
         (\bar{c}_{\beta }  b^{\alpha})_{V-A}    \vspace{0.2cm} \\
Q_2 = &  (\bar{s}_{\alpha}  c^{\alpha})_{V-A}
         (\bar{c}_{\beta }  b^{\beta })_{V-A}     \vspace{0.2cm} \\
Q_3 = &  (\bar{s}_{\alpha}  b^{\alpha})_{V-A}
  \sum_q (\bar{q}_{\beta }  q^{\beta })_{V-A}    \vspace{0.2cm} \\
Q_4 = &  (\bar{s}_{\alpha}  b^{\beta })_{V-A}
  \sum_q (\bar{q}_{\beta }  q^{\alpha})_{V-A}    \vspace{0.2cm} \\
Q_5 = &  (\bar{s}_{\alpha}  b^{\alpha})_{V-A}
  \sum_q (\bar{q}_{\beta }  q^{\beta })_{V+A}     \vspace{0.2cm} \\
Q_6 = &  (\bar{s}_{\alpha}  b^{\beta })_{V-A}
  \sum_q (\bar{q}_{\beta }  q^{\alpha})_{V+A}    \vspace{0.2cm} \\
Q_7 = & \f{e}{8 \pi^2} m_b \bar{s}_{\alpha} \sigma^{\mu \nu}
(1+\gamma_5) b^{\alpha} F_{\mu \nu}            \vspace{0.2cm} \\
Q_8 = & \f{g}{8  \pi^2} m_b \bar{s}_{\alpha} \sigma^{\mu \nu}
(1+\gamma_5) (T^a)^{\alpha}_{\beta} b^{\beta} G_{\mu \nu}^a
\end{array}
\ee

\noindent where $(\bar{q} q')_{V \pm A} = \bar{q} \gamma_{\mu} (1 \pm \gamma_5)
q'$ and $\sigma_{\mu \nu} = \f{i}{2} [\gamma_{\mu},\gamma_{\nu}]$.
$Q_{1,2}$ are the current-current operators, $Q_{3-6}$ the QCD penguin
operators, and $Q_{7,8}$ the ``magnetic penguin'' operators.  In the
latter operators we have neglected the terms proportional to the
s-quark mass since they give only a correction of order $m^2_s/m^2_b$
to the decay rate.

	The inclusion of the couplings in the definition of $Q_7$ and
$Q_8$ allows one to discuss the renormalization group evolution in the
full analogy with the case of the usual $\Delta S =1$ or $\Delta B =1$
hamiltonians for nonleptonic decays.

	Following \cite{BJLW} we write

\be \label{evol}
\vec{C}(\mu) = \hat{U}(\mu,M_W) \vec{C}(M_W)
\ee

\noindent where the renormalization group evolution matrix is given
generally by

\be \label{umatrix1}
\hat{U}(m_1,m_2) = T_g exp \int_{g(m_2)}^{g(m_1)} dg'
\f{\hat{\gamma}^T(g')}{\beta(g')}
\ee

\noindent with $m_1 < m_2$. Here $T_g$ denotes such ordering in the coupling
constants that they increase from right to left.  Next,
$\hat{\gamma}(g)$ is the $8 \times 8$ anomalous dimension matrix of
the operators $Q_i$, and $\beta(g)$ is the usual renormalization group
function which governs the evolution of $\al$.

	Keeping the first two terms in the expansions for $\gamma(g)$
and $\beta(g)$

\be \label{gamma}
\hat{\gamma}(g) = \hat{\gamma}^{(0)} \f{g^2}{16 \pi^2} +
		  \hat{\gamma}^{(1)} \f{g^4}{(16 \pi^2)^2}
\;\; ; \;\;\;\;\;\;\;\;\;\;\;\;\;\;\;\;\;\;\;\;\;
\beta(g) = -\beta_0 \f{g^3}{16 \pi^2} - \beta_1 \f{g^5}{(16 \pi^2)^2},
\ee

\noindent where $\beta_0=11-\f{2}{3}f$ and $\beta_1=102-\f{38}{3}f$,
one finds \cite{BJLW}:

\be \label{umatrix2}
\hat{U}(m_1,m_2) = \left( \hat{1} + \f{\al(m_1)}{4 \pi} \hat{J} \right)
			\hat{U}^{(0)}(m_1,m_2) \left( \hat{1} -
		   \f{\al(m_2)}{4 \pi} \hat{J} \right)
\ee

\noindent where $\hat{U}^{(0)}(m_1,m_2)$ denotes the evolution matrix
in the leading logarithmic approximation and $\hat{J}$ summarizes the
next-to-leading corrections to this evolution. If

\be \label{VG}
\hat{\gamma}^{(0)}_D \equiv \hat{V}^{-1}\hat{\gamma}^{(0)T}\hat{V};
\hspace{3cm}
\hat{G} \equiv \hat{V}^{-1}\hat{\gamma}^{(1)T} \hat{V}
\ee

\noindent where $\hat{\gamma}^{(0)}_D$ denotes a diagonal matrix whose
diagonal elements are the components of the vector
$\vec{\gamma}^{(0)}$, then

\be \label{LLumatrix}
\hat{U}^{(0)}(m_1,m_2) =
\hat{V}
\left[ \left( \f{\al(m_2)}{\al(m_1)} \right)^{\vec{a}} \right]_D
\hat{V}^{-1}
\hspace{0.5cm} {\rm with} \hspace{0.5cm}
\vec{a} = \f{\vec{\gamma}^{(0)}}{2 \beta_0}.
\ee

For the matrix $\hat{J}$ one \vspace{-0.5cm} gets

\be \label{J}
\hat{J} = \hat{V} \hat{S} \hat{V}^{-1}
\ee

\noindent where the elements of $\hat{S}$ are given by

\be \label{S}
S_{ij} = \delta_{ij} \gamma_i^{(0)} \f{\beta_1}{2 \beta_0^2} -
\f{G_{ij}}{2 \beta_0 + \gamma_i^{(0)} - \gamma_j^{(0)}}
\ee

\noindent with $\gamma_i^{(0)}$ denoting the elements of
$\vec{\gamma}^{(0)}$ and $G_{ij}$ the elements of $\hat{G}$.

	The ratio $\al(m_2)/\al(m_1)$ in eq. (\ref{LLumatrix}) has now
to be calculated with help of the two-loop renormalization group
equation for $\al(\mu)$.  Once $\al(M_Z)$ is treated as an initial
condition for this equation, the corresponding solution takes the form

\be \label{alphaNLL}
\al(\mu) = \f{\al(M_Z)}{v(\mu)} \left[1 - \f{\beta_1}{\beta_0}
           \f{\al(M_Z)}{4 \pi}    \f{ln\;v(\mu)}{v(\mu)} \right]
\ee

\noindent where $v(\mu)$ is {\em exactly} as in eq. (\ref{v(mu)}).

	To the same level of accuracy we expand $\vec{C}(M_W)$ as
follows

\be \label{C}
\vec{C}(M_W) = \vec{C}^{(0)}(M_W)
		+ \f{\al(M_W)}{4 \pi} \vec{C}^{(1)}(M_W).
\ee

\noindent The nonvanishing $C^{(0)}_i(M_W)$ have been given in
eqs. (\ref{c2})--(\ref{c8}).

	The leading logarithmic approximation to $C_i(\mu)$ is
completely known. It is obtained by setting $\hat{J}=0$ and
$\vec{C}^{(1)} =0$ in eqs. (\ref{umatrix2}) and (\ref{C})
respectively. In the appendix, we explicitly give the matrix
$\hat{\gamma}^{(0)}$ calculated \cite{Ciu} in a certain regularization
scheme called the HV scheme. The scheme dependence of
$\hat{\gamma}^{(0)}$ will be discussed below. Here we would like only
to remark that the $6 \times 6$ submatrix of $\hat{\gamma}^{(0)}$
describing mixing of $(Q_1,Q_2,...,Q_6)$ and the $2 \times 2$
submatrix for $(Q_7,Q_8)$ follow from one-loop calculations. On the
other hand, the leading order mixing between these two sectors results
from two-loop diagrams.

\ \\ Let us enumerate what has been already calculated in the
literature and which calculations are still required in order to
complete the next-to-leading calculation of $C_i(\mu)$.  Beyond the
leading order, $\hat{\gamma}^{(1)}$ and $\vec{C}^{(1)}(M_W)$ have to
be calculated. Actually, the $6 \times 6$ submatrix of
$\hat{\gamma}^{(1)}$ describing the two-loop mixing of $(Q_1,...,
Q_6)$ and the corresponding initial conditions in $\vec{C}^{(1)}(M_W)$
have been already calculated in refs. \cite{BJLW,page14}. The remaining
ingredients of a next-to-leading analysis of $\vec{C}(\mu)$ are:\\
\ \\
(i) The two-loop mixing in the $(Q_7,Q_8)$ sector of
$\hat{\gamma}^{(1)}$. This calculation is currently being performed
\cite{prepar}.\\
\ \\
(ii) The three-loop mixing between the sectors $(Q_1,...,Q_6)$ and
$(Q_7,Q_8)$ which with our normalizations contribute to
$\hat{\gamma}^{(1)}$.\\
\ \\
(iii) The ${\cal O}(\al)$ corrections to $C_7(M_W)$ and $C_8(M_W)$ in
eqs. (\ref{c7}) and (\ref{c8}). This requires evaluation of two-loop
penguin diagrams with internal W and top quark masses and a proper
matching with the effective five-quark theory. An attempt to calculate
the necessary two-loop Standard Model diagrams has been recently made
in ref. \cite{Yao2}. The effective theory one-loop diagrams with $Q_7$
and $Q_8$ insertions have been calculated in ref. \cite{AG2}. However,
the finite parts of the effective theory two-loop diagrams with the
insertions of the four-quark operators (see figs. 5 and 6 of ref.
\cite{MMNP}) are still unknown.\\
\ \\
\indent All these calculations are very involved, and the necessary
three-loop calculation is a truly formidable task! Yet, as will be
evident from the following discussion, all these calculations have to
be done if we want to reduce the theoretical uncertainties in \bsg to
around 10\%.

	It is important to stress that a next-to-leading calculation
performed without resumming large logarithms $ln(M^2_W/m^2_b)$ would
not be more accurate than the present leading-order one where these
logarithms are resummed. This is why the calculation of the three-loop
mixing is unavoidable.\\
\ \\
\indent Once the evolution of the coefficients $C_i(\mu)$ down to the
scale $\mu \sim m_b$ is performed, one has to evaluate the matrix
elements of all the operators at this scale.

	In the leading order, only the tree-level matrix element of
$Q_7$ and the one-loop matrix elements of $Q_1-Q_6$ that are of order
$\al^0$ have to be included. The latter matrix elements vanish for the
on-shell photon in any 4-dimensional regularization scheme and also in
the HV scheme, i.e. dimensional regularization with the
't~Hooft-Veltman definition of $\gamma_5$ \cite{HV}. All but the last
\cite{Ciu} calculations of the leading order QCD corrections to \bsg
used the NDR scheme, i.e. the dimensional regularization with fully
anticommuting $\gamma_5$.\footnote{ A complete calculation using the
``dimensional reduction'' scheme is not yet existent - see
ref. \cite{MMDRED}.} In this scheme, the one-loop \bsg matrix elements
of some of the four-quark operators $Q_1,...,Q_6$ do not vanish for
the on-shell photon but are proportional to the tree-level matrix
element of $Q_7$ (even for the quarks off-shell).

	The regularization scheme-dependence of leading order matrix
elements is a peculiar feature of the \bsg decay. It can arise because
one-loop mixing between the $(Q_1,...,Q_6)$ and $(Q_7,Q_8)$ sectors
vanishes. In consequence, what usually would be a next-to-leading
order effect is only a leading one. And what would usually be
next-next-to-leading (like the above-mentioned three-loop mixing) is
only next-to-leading.

	This peculiarity causes, that at the leading order it is
convenient to introduce the so-called ``effective coefficients'' for
the operators $Q_7$ and $Q_8$: First, we observe that in any
regularization scheme one can write the one-loop matrix elements of
the four-quark operators $Q_1,...,Q_6$ as
\be \label{def_y}
\me{ Q_i {}}_{\scs one\;\;loop} = y_i \me{ Q_7}_{\scs tree},
\hspace{1cm} i=1,...,6
\ee

\noindent for the on-shell photon but quarks possibly off-shell. In
consequence, the leading order \bsg matrix element of $H_{eff}$ is
equal to the tree level matrix element of

\be \label{treematrix}
-V_{tb} V^*_{ts} \f{G_F}{\sqrt{2}} C^{(0)eff}_7(\mu) O_7
\ee
\noindent \vspace{-0.5cm} where

\be \label{def_c7eff}
C^{(0)eff}_7(\mu) = C^{(0)}_7(\mu) + \sum_{i=1}^6 y_i C^{(0)}_i(\mu).
\ee

In the HV scheme all the $y_i$'s vanish, while in the NDR scheme
$\vec{y} = (0,0,0,0,-\f{1}{3},-1)$.  This regularization scheme
dependence is cancelled by a corresponding regularization scheme
dependence in $\hat{\gamma}^{(0)}$ (see ref. \cite{Ciu} for
details). Consequently, the quantity $C^{(0)eff}_7(\mu)$ defined in
eq. (\ref{def_c7eff}) and used in eq. (\ref{main}) is
scheme-independent.

	The numbers $y_i$ in eq. (\ref{def_y}) come from divergent,
i.e. purely short-distance parts of the one-loop integrals. So no
reference to the spectator-model is necessary here. This is why one is
allowed to treat the expression (\ref{treematrix}) as the proper
effective hamiltonian that has to be inserted in between hadronic
states in a calculation of, say, the $B \ra K^* \gamma$ decay.

	An ``effective coefficient'' for the operator $Q_8$ introduced
in a similar way equals to

\be
C^{(0)eff}_8(\mu) = C^{(0)}_8(\mu) + \sum_{i=1}^6 z_i C^{(0)}_i(\mu)
\ee

\noindent where the numbers $z_i$ are defined by the one-loop
$b \ra s \; gluon$ on-shell matrix elements of the four-quark
operators:
\be
\me{Q_i}_{\scs one\;\;loop} = z_i \me{Q_8}_{\scs tree},
\hspace{1cm} i=1,...,6.
\ee

\noindent All the $z_i$'s vanish in the HV scheme, while in the NDR
scheme we have $\vec{z} = (0,0,0,0,1,0)$.

	For later convenience we introduce a scheme-independent
vector

\be
\vec{C}^{(0)eff}(\mu) = \left[ C^{(0)}_1(\mu),...., C^{(0)}_6(\mu),
C^{(0)eff}_7(\mu),C^{(0)eff}_8(\mu) \right] .
\ee
\noindent From the RGE for $\vec{C}^{(0)}(\mu)$ it is straighforward
to derive the RGE for $\vec{C}^{(0)eff}(\mu)$. It has the form
\be \label{RGEeff}
\mu \f{d}{d \mu} C^{(0)eff}_i(\mu) = \gamma^{(0)eff}_{ji} C^{(0)eff}_j(\mu)
\ee
\noindent \vspace{-0.5cm} where

\be \label{def.geff}
\gamma^{(0)eff}_{ji} = \left\{ \begin{array}{cc}
\gamma^{(0)}_{j7} +
\sum_{k=1}^6 y_k\gamma^{(0)}_{jk} -y_j\gamma^{(0)}_{77} -z_j\gamma^{(0)}_{87}
& \mbox{when $i=7$ and $j=1,...,6$}\\
\gamma^{(0)}_{j8} +
\sum_{k=1}^6 z_k\gamma^{(0)}_{jk} -z_j\gamma^{(0)}_{88}
& \mbox{when $i=8$ and $j=1,...,6$}\\
\gamma^{(0)}_{ji} & \mbox{otherwise.}
\end{array}
\right.
\ee

\noindent The matrix $\hat{\gamma}^{(0)eff}$ is a scheme-independent quantity.
We give it explicitly in the appendix.

	At the next-to-leading level the situation is not as simple.
One has then to include one-loop \bsg matrix elements of $Q_7$ and
$Q_8$, and two-loop matrix elements of $Q_1,...,Q_6$. As long as the
photon and the external quarks are put on-shell, we can still formally
write the considered matrix elements as proportional to the tree-level
matrix element of $Q_7$. One could naively expect that the
proportionality constants depend only on the ratio $\mu/m_b$ because
after neglection of the s-quark mass, we have only a single scale in
the process.  Unfortunately, they also depend on the infrared
regulator used in the calculation.\footnote{The infrared divergencies
are removed later by including soft gluon bremsstrahlung - see
ref. \cite{AG2}.} However, their $\mu$-dependence is still a purely
short-distance effect and can be found with help of the existing
leading-order results.

	As discussed in ref. \cite{BJL}, the $\mu$-dependence of
$\me{\vec{Q}(\mu)}$ can be calculated in a renormalization group
improved perturbation theory, as long as $\mu$ is not too small. Taking
the scale $m_b$ as a reference point, the $\mu$-dependence of
$\me{\vec{Q}(\mu)}$ is simply given by

\be \label{m.elem}
\me{\vec{Q}^T(\mu)} = \me{\vec{Q}^T(m_b)} \hat{U}(m_b,\mu)
\ee

\noindent with $\hat{U}$ given already in eq. (\ref{umatrix2}). The
scale $m_b$ is chosen here only for convenience. Inserting
eqs. (\ref{evol}) and (\ref{m.elem}) into the final expression for the
\bsg amplitude:
\be \label{amplitude}
A(b \ra s \gamma) = -V_{tb} V^*_{ts} \f{G_F}{\sqrt{2}}
				\sum_{i=1}^{8} \me{Q_i(\mu)} C_i(\mu)
\ee
\noindent one finds that the $\mu$-dependence present in $\vec{C}(\mu)$ is
cancelled by the corresponding $\mu$-dependence in $\me{\vec{Q}(\mu)}$.

	Let us now write the sum of the tree and one-loop on-shell
matrix elements of $Q_7$ and $Q_8$ calculated at $\mu = m_b$ as

\be \label{def1.ri7}
\me{Q_i(m_b)} = \left[ \delta_{i7} + \f{\al(m_b)}{4\pi} r_{i7} \right]
\me{Q_7(m_b)}_{tree},\;\;\;\;\;\;\;i=7,8,
\ee

\noindent and the sum of one- and two-loop on-shell matrix elements of $Q_1,
..., Q_6$ as

\be \label{def2.ri7}
\me{Q_i(m_b)} = \left [ y_i + \f{\al(m_b)}{4 \pi} r_{i7} \right]
\me{Q_7(m_b)}_{tree},\;\;\;\;\;\;\;i=1,...,6.
\ee

\noindent As long as the infrared divergencies are regulated
dimensionally, the $r_{i7}$'s do not contain dimensionful parameters,
i.e. they are either pure numbers or depend only on the infrared
regulator $\epsilon_{\scs IR}$.

	If $\mu \sim m_b$, there are no large logarithms present in
$\me{\vec{Q}(\mu)}$, and the resummation of logarithms is not needed
in eq. (\ref{m.elem}). Therefore expanding eq. (\ref{m.elem}) around
$\mu = m_b$ we can reproduce the $\mu$-dependent terms that would
appear in the straightforward perturbative calculation of
$\me{\vec{Q}(\mu)}$. Using eqs. (\ref{m.elem})--(\ref{def2.ri7}) and
expanding the matrix $\hat{U}(m_b,\mu)$ from eq. (\ref{m.elem}) around
$\mu=m_b$ we find, that the \bsg amplitude calculated in terms of the
coefficients renormalized at any scale $\mu \sim m_b$ can be written
in the following form:

\be     \label{NLLelement}
A(b \ra s \gamma) = -V_{tb} V^*_{ts} \f{G_F}{\sqrt{2}}\;\; D\;\;
\me{Q_7(m_b)}_{tree}
\ee
\noindent where
\be    \label{def_D}
D = C_7^{eff}(\mu) + \f{\al(m_b)}{4 \pi}
\left( C_i^{(0)eff}(\mu)  \gamma^{(0)eff}_{i7} ln \f{m_b}{\mu}
+ C_i^{(0)}(\mu) r_{i7} \right).
\ee

\noindent All the $C_i$'s in the expression (\ref{def_c7eff}) for
$C_7^{eff}(\mu)$ are now taken up to next-to-leading
accuracy.\footnote{ This refers to the first term in D. In the term
proportional to $\al$ one can still keep the leading logarithmic
$C^{(0)eff}_7$. Note, that $r_{i7}$'s are multiplied by the usual, not
the effective coefficients. The resulting scheme-dependences should
cancel with the ones present in $r_{i7}$ and $\vec{C}^{(1)}(\mu)$.}
The term proportional to $ln \f{m_b}{\mu} $ comes solely from
expanding $\me{\vec{Q}(\mu)}$ around $\mu=m_b$. The tree-level matrix
element of $Q_7$ is denoted as $\me{Q_7(m_b)}_{tree}$ in order to
point out that the mass $m_b$ in the normalization of this operator is
renormalized at the scale $\mu =m_b$.

	A next-to-leading generalization of eq. (\ref{main}) takes the
following form:

\be                   \label{NLLmain}
R =
\f{\Gamma[b \ra s \gamma]}{\Gamma[b \ra c e \bar{\nu}_e]}
 =  \f{|V_{ts}^* V_{tb}|^2}{|V_{cb}|^2}
\f{6 \alpha_{\scs QED}}{\pi g(z)} F |D|^2.
\ee

with D defined in eq. (\ref{def_D}), and the factor F given by

\be \label{factor}
F = \left( \f{m_b(\mu=m_b)}{m_b^{pole}} \right)^2 \f{1}{\kappa(z)}.
\ee

\noindent The quantity $\kappa(z)$ is a sizable next-to-leading QCD
correction to the semileptonic decay \cite{kappa}

\be  \label{kappa}
\kappa(z) \simeq 1 - \frac{2 \al (m_b)}{3 \pi}
\left[ (\pi^2 -\frac{31}{4})(1-z)^2+\frac{3}{2} \right].
\ee

\noindent which tends to cancel with the other factor in eq.  (\ref{factor})

\be    \label{mass.ratio}
\left( \f{m_b(\mu=m_b)}{m_b^{pole}} \right)^2  =
1 - \f{8}{3} \f{\al(m_b)}{\pi},
\ee

\noindent and, in consequence, the expression (\ref{factor}) differs
from unity by only around 8\%.

	The origin of the latter factor is very simple: We have
normalized the \bsg rate to the semileptonic rate in order to cancel
factors $m_b^5$ in front of the spectator-model expressions. However,
in the semileptonic case, all five factors of $m_b$ come from the
on-shell external lines, while in the case of \bsg two of the $m_b$'s
come from the normalization of $Q_7$, and should be taken in the same
renormalization scheme as the coefficients. This becomes even more
obvious when one realizes that the anomalous dimension matrix element
$\gamma^{(0)}_{77} = \f{32}{3}$ is dominated by the anomalous
dimension $\gamma^{(0)}_m = \f{24}{3} = 8$ of the mass $m_b$ which
normalizes $Q_7$. If we did not include the running $m_b(\mu)$ in the
operator normalization, then our coefficients would run differently
than described in eq. (\ref{c7eff}).

	Let us now return to the complete next-to-leading expressions
(\ref{NLLelement})--(\ref{def_D}) for the \bsg amplitude. From
eq. (\ref{def_D}) and from the RGE (\ref{RGEeff}) for
$\vec{C}^{(0)eff}(\mu)$ one can easily see that the dominant
$\mu$-dependence in the leading logarithmic coefficient
$C^{(0)eff}_7(\mu)$ will be cancelled by the logarithms coming from
the next-to-leading matrix elements. Consequently, the resulting
$\mu$-dependence of the obtained result will be one order higher than
the accuracy of the performed calculation, as it always must happen in
any perturbative calculation.

	We can explicitly check the cancellation of $\mu$-dependence,
because the coefficients at the logarithms in eq. (\ref{def_D})
are given by the matrix $\hat{\gamma}^{(0)eff}$ which is already known
(see the appendix). However, any change of $\mu$ is equivalent to a
shift in the unknown constant terms $r_{i7}$. Consequently, a
meaningful analysis of the $\mu$-dependence must also include
$r_{i7}$. These constants are generally renormalization-scheme
dependent, and this dependence can only be cancelled by calculating
$\hat{\gamma}^{(1)}$ in the same renormalization scheme. Since this
point has been extensively discussed in several papers
\cite{BJLW,page14,BJL}, we will not repeat this discussion here. However,
it is clear from these remarks, that in order to address the
$\mu$-dependence and the renormalization-scheme dependence as well as
their cancellations, it is necessary to perform a complete
next-to-leading order analysis of $\vec{C}(\mu)$ along the lines
presented above, and to calculate the constants $r_{i7}$ in
eqs. (\ref{def1.ri7}) and (\ref{def2.ri7}).

	The program of calculating the complete next-to-leading
contributions and reducing the $\mu$-dependence has been already
accomplished in several other weak decays. In ref. \cite{A1} the
$\mu$-dependence in top-quark contributions to $K^0$-$\bar{K}^0$ and
$B^0$-$\bar{B}^0$ mixings have been considered. A similar analysis of
the charm quark contribution to $K^0$-$\bar{K}^0$ mixing has been
recently done in ref. \cite{A2}. Next, in ref. \cite{A3} the
$\mu$-dependences in rare decays $K_L \ra \pi^0 \nu \bar{\nu}$, $K^+
\ra \pi^+ \nu \bar{\nu}$, $K_L \ra \mu \bar{\mu}$, $B \ra
l \bar{l}$ and $B \ra X_s \nu \bar{\nu}$ have been studied.  In all
these papers it has been demonstrated explicitly that the inclusion of
next-to-leading corrections considerably reduced the $\mu$-dependence
present in the leading order expressions. The remaining
$\mu$-dependence can be further reduced only by including still higher
order terms in the renormalization group improved perturbation theory.

	In view of the tremendous complexity of the next-to-leading
calculations for $b \ra s \gamma$, we are not yet in a position to
make a similar analysis for this decay. We can however perform a
simple exercise with the terms we already know, in order to get a
rough idea by how much a future next-to-leading order calculation
could reduce the $\mu$-dependence shown by the solid line in
fig. \ref{mudep}.

	We proceed as follows: We set $\hat{J}=0$ and
$\vec{C}^{(1)}=0$ in eqs. (\ref{umatrix2}) and (\ref{C})
respectively, i.e. we calculate $\vec{C}^{eff}(\mu)$ in the
leading logarithmic approximation. $C^{(0)eff}_7(\mu)$ has been
already given in eq. (\ref{c7eff}). For the remaining coefficients we
find
\begin{eqnarray}
\begin{array}{ccc}
\vspace{0.2cm}
C^{(0)}_1(\mu)=& \hspace{-0.4cm}
		 -\f{1}{2} \eta^{-\f{12}{23}} + \f{1}{2} \eta^{\f{6}{23}}&\\
\vspace{0.2cm}
C^{(0)}_2(\mu)=& \hspace{-0.1cm}
                  \f{1}{2} \eta^{-\f{12}{23}} + \f{1}{2} \eta^{\f{6}{23}}&\\
\vspace{0.2cm}
C^{(0)}_3(\mu)=& \hspace{-0.1cm}
                  \f{1}{6} \eta^{-\f{12}{23}} - \f{1}{14} \eta^{\f{6}{23}}&
\hspace{-0.4cm} + 0.0510 \eta^{0.4086}   - 0.1403 \eta^{-0.4230}
                - 0.0113 \eta^{-0.8994}  + 0.0054 \eta^{0.1456}\\
\vspace{0.2cm}
C^{(0)}_4(\mu)=& \hspace{-0.4cm}
                 -\f{1}{6} \eta^{-\f{12}{23}} - \f{1}{14} \eta^{\f{6}{23}}&
\hspace{-0.4cm} + 0.0984 \eta^{0.4086}   + 0.1214 \eta^{-0.4230}
                + 0.0156 \eta^{-0.8994}  + 0.0026 \eta^{0.1456}\\
\vspace{0.2cm}
C^{(0)}_5(\mu)=&&\hspace{-0.4cm}
                -0.0397 \eta^{0.4086}   + 0.0117 \eta^{-0.4230}
                -0.0025 \eta^{-0.8994}   + 0.0304 \eta^{0.1456}\\
\vspace{0.2cm}
C^{(0)}_6(\mu)=&& \hspace{-0.2cm}
		  0.0335 \eta^{0.4086}   + 0.0239 \eta^{-0.4230}
                 -0.0462 \eta^{-0.8994}  - 0.0112 \eta^{0.1456}
\end{array} \nonumber \end{eqnarray}
and
\begin{eqnarray}
C^{(0)eff}_8(\mu) = (C^{(0)}_8(M_W)  + \f{313063}{363036}) \eta^{\f{14}{23}}
- \hspace{10cm}
\nonumber \end{eqnarray} \vspace*{-0.55cm}
\begin{eqnarray} \hspace{4.5cm}
              -0.9135 \eta^{0.4086}   +  0.0873 \eta^{-0.4230}
              -0.0571 \eta^{-0.8994}  -  0.0209 \eta^{0.1456}
\nonumber \end{eqnarray}

\noindent Next, we set\footnote{
Here we assume we have already removed the infrared divergencies from
all the $r_{i7}$'s by adding soft gluon bremsstrahlung.}
$r_{i7}=0$ in eq. (\ref{NLLelement}), except
for $r_{27}$ for which we consider values $-2$, $0$ and $2$. It is
clear from eqs. (\ref{def1.ri7})--(\ref{def2.ri7}) that all the
$r_{i7}$'s can take values of order 1. We choose to vary $r_{27}$
because it is the operator $Q_2$ that has the largest coefficient ---
the coefficient $C_2^{(0)}(m_b)$ is around 1.1. The
``next-to-leading'' results for the ratio R found this way are
presented as functions of $\mu$ by the dashed lines in
fig. \ref{mudep}.

	One can see that the $\mu$-dependence is much weaker than for
the leading-order results (the solid line), but not completely
absent. The initial $\pm 25\%$ uncertainty is reduced to around $\pm
3\%$.  The dominant $\mu$-dependence caused by factors of order
$\al(m_b) ln(m_b/\mu)$ cancels out, but we still have noncancelling
higher order terms like $\al^2(m_b) ln^2(m_b/\mu)$ or $\al^2(m_b)
ln(m_b/\mu)$. We have organized our exercise in such a way that both
kinds of the latter terms do appear, in spite of that we have set a
lot of the next-to-leading corrections to zero. The higher order
$\mu$-dependence could be, in principle, removed by including still
higher order QCD corrections.

	Now it is easy to see that, indeed, a reasonable variation in
possible non-logarithmic next-to-leading terms, namely changing
$r_{27}$ from $-2$ to $2$, causes a similar change in the final
results as varying $\mu$ by a factor of 2 in the leading order
terms. This supports the method we have used for estimating the size
of the NLL corrections in our phenomenological discussion in section
3.

	The simple exercise presented above explicitly illustrates the
cancellation of the $\mu$-depen- dence present in the leading-order
expressions. It cannot, however, replace the full next-to-leading
order calculation. Indeed, we still do not know which of the dashed
curves in fig. \ref{mudep} is closest to the true next-to-leading
result. Our exercise is, nevertheless, rather encouraging as it shows
that the theoretical prediction for \Bsg will be considerably improved
once all the next-to-leading calculations are completed.\\
\ \\
\noindent {\bf 5. Conclusions}\\
\ \\
\indent i) Among many theoretical uncertainties present in the calculation of
the \Bsg decay, the most important one is due to the absence of a
complete calculation of next-to-leading short-distance QCD corrections
to this decay. It causes around $\pm 25\%$ inaccuracy in the prediction
for this decay rate, while all the other uncertainties are around or
below 10\%.

ii) Once the present experimental data are taken into account, it is
quite improbable that a future precise measurement of the \Bsg rate
will contradict the Standard Model prediction, in view of its large
uncertainty. However, if the future measurement of the inclusive rate
is close to the present upper bound, then some useful correlation
between $m_t$ and $\al(M_Z)$ will emerge.

iii) In spite of the theoretical uncertainties, the present
experimental data still put quite strong bounds on the parameters of
the Two-Higgs-Doublet Model. The $b \ra s \gamma$, $Z \ra b \bar{b}$
and $b \ra c \tau \bar{\nu}_{\tau}$ processes are enough to give the
most stringent bounds in the $M_{H^{\pm}}$-$tan \beta$ plane. However,
our \bsg bounds are weaker than found by other authors.

iv) The theoretical prediction can be significantly improved by making
a systematic next-to-leading calculation within the framework of
renormalization group improved perturbation theory. Such a calculation
is very involved, since it requires a calculation of finite parts of
many two-loop diagrams, and divergent parts of three-loop diagrams.
However, it has to be done if we want to make the uncertainty coming
from short-distance effects smaller than the other uncertainties.

v) The uncertainty in the leading order prediction is best signaled by
its strong dependence on the renormalization scale $\mu$ at which the
RGE evolution is terminated. We have described in detail how this
dependence is reduced in a full next-to-leading result, and performed
a simple exercise to check it numerically. As a result, we can
conclude that once a next-to-leading result is found, its
$\mu$-dependence should be below 5\%, for $\mu$ changing from $m_b/2$
to $2m_b$.

\ \\
\noindent {\bf Acknowledgements}\\

	We would like to thank the authors of refs. \cite{Hollik,Park}
for sending their results to us prior to publication.

\ \\
\noindent {\bf 6. Appendix}\\
\ \\
\indent Here we give the scheme-independent matrix
$\hat{\gamma}^{(0)eff}$ defined in eq. (\ref{def.geff}) that governs
the leading-logarithmic QCD corrections to \bsg. In the HV scheme this
matrix is equal to the matrix $\hat{\gamma}^{(0)}$ of eq. (\ref{gamma}).

\be
\hat{\gamma}^{(0)eff} = \left[
\begin{array}{cccccccc}
\vspace{0.2cm}
-2 & 6 &     0     &      0     &     0    &       0     &            0
    &       3       \\
\vspace{0.2cm}
 6 &-2 &-\f{2}{9}  &  \f{2}{3}  &-\f{2}{9} &  \f{2}{3}   &       \f{416}{81}
    &  \f{70}{27}   \\
\vspace{0.2cm}
 0 & 0 &-\f{22}{9} & \f{22}{3}  &-\f{4}{9} &  \f{4}{3}   &      -\f{464}{81}
    &\f{140}{27}+3f \\
\vspace{0.2cm}
 0 & 0 &6-\f{2}{9}f&-2+\f{2}{3}f&-\f{2}{9}f&  \f{2}{3}f  &
\f{416}{81}u-\f{232}{81}d& 6+\f{70}{27}f \\
\vspace{0.2cm}
 0 & 0 &     0     &      0     &     2    &      -6     &        \f{32}{9}
    & -\f{14}{3}-3f \\
\vspace{0.2cm}
 0 & 0 &-\f{2}{9}f & \f{2}{3}f
&-\f{2}{9}f&-16+\f{2}{3}f&-\f{448}{81}u+\f{200}{81}d&-4-\f{119}{27}f\\
\vspace{0.2cm}
 0 & 0 &     0     &      0     &     0    &       0     &        \f{32}{3}
    &       0       \\
\vspace{0.2cm}
 0 & 0 &     0     &      0     &     0    &       0     &       -\f{32}{9}
    &   \f{28}{3}   \\
\end{array} \right] \nonumber \ee

\noindent where f = u + d, and u (d) is the number of up- (down-) quarks
in the effective theory. In our case $u=2$ and $d=3$.

	The above matrix has been obtained under the assumption that
the disagreements between refs. \cite{MMNP}, \cite{Yao1} and
\cite{Ciu} concerning $\gamma^{(0)}_{57}$, $\gamma^{(0)}_{58}$ and
$\gamma^{(0)}_{68}$ in the NDR scheme are resolved in favour of ref.
\cite{Ciu}. More information about these disagreements is contained in
the appendix of ref. \cite{MMDRED}. Here we only mention that these
disagreements make the numerical leading order results differ by only
around 1\%, so they are irrelevant for our phenomenological
discussion. Note, that our normalization of $Q_7$ and $Q_8$ differs
from the one used in ref. \cite{Ciu}.

	The scheme-independent numbers $a_i$ and $h_i$ in
eq. (\ref{c7eff}) are given by the eigenvalues and eigenvectors of
$\hat{\gamma}^{(0)eff}$ .\footnote{according to eqs. (\ref{VG}) and
(\ref{LLumatrix}) with $\hat{\gamma}^{(0)}$ replaced by
$\hat{\gamma}^{(0)eff}$} Their explicit values are given by

\be
\begin{array}{ccccccccc}
\vspace{0.2cm}
a_i = ( &    \f{14}{23},     &     \f{16}{23},   & \f{6}{23},&-\f{12}{23},
        &      0.4086,       &      -0.4230,     & -0.8994,  &   0.1456     )
\\
h_i = ( &\f{626126}{272277}, &-\f{56281}{51730}, &-\f{3}{7}, & -\f{1}{14},
        &     -0.6494,       &      -0.0380,     & -0.0186,  &  -0.0057     )
\end{array}
\ee

With help of {\it Mathematica} \cite{Wolfram}, all the above numbers can
be found exactly. However, some of them are so complicated that it is
much better to keep them numerically.

\end{document}